\documentclass[9pt,conference, compsocconf]{sig-alternate}
\usepackage{url}
\usepackage{cite}
\usepackage{balance}
\usepackage[vlined,figure]{algorithm2e}
\usepackage{amssymb}
\usepackage{amsmath}
\usepackage{booktabs}
\usepackage{color}
\usepackage{graphicx}
\usepackage{wrapfig}
\usepackage{tikz}

\usepackage{listings}
\newcommand\lt[1]{{\lstinline+#1+}}
\definecolor{darkblue}{rgb}{0,0,0.5}
\definecolor{dkgreen}{rgb}{0,0.5,0}
\definecolor{dkred}{rgb}{0.5,0,0}
\definecolor{gray}{rgb}{0.5,0.5,0.5}
\definecolor{dkgray}{rgb}{0.3,0.3,0.3}
\usepackage[capitalize]{cleveref}

\newcommand{\oomit}[1]{}

\conferenceinfo{ICSE}{'14, May 31 -- June 7, 2014, Hyderabad, India}
\CopyrightYear{14}
\crdata{978-1-4503-2756-5/14/05}

\begin{document}

\title{Using Dynamic Analysis to Generate Disjunctive Invariants}

\numberofauthors{4}
\author{
\alignauthor
ThanhVu Nguyen\\
       \affaddr{Computer Science}\\
       \affaddr{University of New Mexico}\\
       \affaddr{New Mexico, USA}
       \email{tnguyen@cs.unm.edu}
\alignauthor
Deepak Kapur\\
       \affaddr{Computer Science}\\
       \affaddr{University of New Mexico}\\
       \affaddr{New Mexico, USA}
       \email{kapur@cs.unm.edu}
\and
\alignauthor
Westley Weimer\\
       \affaddr{Computer Science}\\
       \affaddr{University of Virginia}\\
       \affaddr{Virginia, USA}
       \email{weimer@cs.virginia.edu}
\alignauthor
Stephanie Forrest
       \affaddr{Computer Science}\\
       \affaddr{University of New Mexico}\\
       \affaddr{New Mexico, USA}
       \email{forrest@cs.unm.edu}
}

\maketitle

\begin{abstract}
  Program invariants are important for defect detection, program
  verification, and program repair. However, existing techniques
  have limited support for important classes of invariants such as disjunctions,
  which express the semantics of conditional statements.  We propose a
  method for generating disjunctive invariants over numerical domains, which   %VU219: added over the numerical domains for clarification
  are inexpressible using classical convex polyhedra.  Using dynamic
  analysis and reformulating the problem in non-standard ``max-plus''
  and ``min-plus'' algebras, our method constructs hulls over program
  trace points.
  Critically, we introduce and infer a weak class of
  such invariants that balances expressive power against
  the computational cost of generating nonconvex shapes in high
  dimensions.

  Existing dynamic inference techniques often generate spurious
  invariants that fit some program traces but do not generalize.  With
  the insight that generating dynamic invariants is easy, we propose
  to verify these invariants statically using $k$-inductive SMT
  theorem proving which allows us to validate invariants that are not
  classically inductive.

Results on difficult kernels involving 
nonlinear arithmetic and abstract arrays suggest that this hybrid approach
efficiently generates and proves correct program invariants.

\iffalse
Building on our earlier work on using traces to generate
geometric shapes based invariants, this paper extends the
approach to consider nonconvex geometric shapes based invariants
of programs using dynamic analysis. 
We build convex hulls for max and min-plus polyhedra over program trace points to represent forms of disjunctive polynomial invariants whose shapes are not expressible using classical convex polyhedra.
To deal with high complexity
of generating nonconvex shapes in high dimensions, 
we introduce a simpler form of max and min-plus relations that balances expressive power with computation cost
Since dynamic analysis is prone to
generating spurious invariants, DIG has been integrated with a
k-induction theorem prover using modern fast SMT technology
to filter out spurious formulas alleged as invariants by dynamic
analysis as well as verifying correct invariants generated. This
integration is based on the view that invariants are difficult
and expensive to automatically generate using static analysis but relative
easier to verify using SMT technology.
Preliminary results show that our hybrid approach can efficiently
generate sound program invariants, i.e., using dynamic analysis to
quickly find complex candidate invariants from program traces and static
analysis to formally verify these results from the program code.
\fi

\end{abstract}




\section{Introduction}

Program invariants are logical properties that hold at certain program
locations. Invariants are important for defect detection
(e.g.,~\cite{slam,blast,ESP}), program verification
(e.g.,~\cite{demoura:tacas08:z3,CCF05,leroy:popl06:}), and even program repair
(e.g.,~\cite{zeller2010,weimer06}).  Invariants can be found using
static or dynamic program analyses.  Static reasoning about source
code can generate invariants without executing the program, but
is often expensive and therefore considers relatively simple forms of
invariants.  In contrast, dynamic analyses infer invariants from
execution traces~\cite{daikon}.  The quality and completeness of these
traces determine the accuracy of the inferred invariants.  As a
result, dynamic analyses often produce spurious invariants that match
some observations but are not sound with respect to general program
behavior. However, dynamic analyses are generally more efficient and
can be targeted to discover more complex forms of invariants.

Existing invariant inference techniques tend to focus on conjunctive,
polynomial and convex invariants. Polynomial invariants, which are
relations among polynomials over numerical program variables, are
particularly important for many applications. As one example,
polynomial inequalities are used to represent pointer arithmetic and other
memory related properties~\cite{CCF05}.  Inspired by abstract
interpretation approaches in static analysis~\cite{cousot:77:abstract,CH78}, recent
dynamic analysis methods use geometric shapes to represent polynomial
invariants~\cite{nguyen:icse12:dig,nguyen:tosem14:dig}. 
Although these convex shapes capture conjunctions of polynomial
relations, they cannot represent disjunctive program properties.

Disjunctive invariants, which represent the semantics of branching, are
more difficult to analyze but crucial to many
programs. For example, after {\lt if (p) \{a=1;\} else \{a=2;\}} 
neither $a=1$ nor $a=2$ is an invariant, but $(p \wedge a=1) \vee (\neg p
\wedge a=2)$ is a disjunctive invariant. Disjunctive invariants thus
capture path-sensitive reasoning, such as those found in most
sorting and searching tasks, as well as functions like \lt{strncpy} in the C standard library.

Existing approaches thus suffer from the twin problems of soundness and
expressive power: Sound static approaches are too inefficient to target
complex and expressive invariants, while efficient dynamic approaches often
yield spurious invariants. For example,  Interproc~\cite{jeannet2010interproc}, a popular static analyzer that employs different abstract domains, and Astr$\acute{\text{e}}$e~\cite{CCF05,BCC03}, a successful program analyzer used for verifying the absence of run-time errors in Airbus avionic systems,
consider only conjunctive invariants, and thus lack expressive power. 
Dynamic convex hull methods capture complex structures but can yield many %VU219: remove the word "too"
spurious invariants.  In fact, such approaches are not used by default
because of false positive issues~\cite{nguyen:tosem14:dig}, and instead users are
asked to specify invariant shapes manually.

We address both expressive power and soundness with a hybrid technique
combining a novel method for inferring expressive invariants
dynamically with a static approach for validating invariants by formal
proof.  At the heart of our dynamic analysis is the insight that
disjunctive invariants, which are not classically convex, can be
reformulated in a non-standard algebra.
Once
reformulated, inference proceeds using a variant of existing geometric
hull approaches. Our static verification technique rests on the
observation that many practical program invariants are $k$-inductive
but not classically
inductive~\cite{sheeran2000checking,KahTin-PDMC-11,donaldson11}. That
is, they can be proved by considering $k$ base cases with an inductive
step that has access to the $k$ previous instances. Our hybrid
algorithm leverages the fact that it is easier to infer complex
candidate invariants dynamically and verify them statically.

We build convex hulls for a special type of nonconvex polyhedra called
\emph{max-plus} to capture certain disjunctive information.  A
polyhedron using max-plus algebra
is a set of relations of the form
$\max(c_0,c_1+v_1,\dots,c_n+v_n) \ge \max(d_0,d_1+v_1,\dots,d_n+v_n)$  over
program variables $v_i$ with coefficients $c_i,d_i \in
\mathbb{R}\cup\{-\infty\}$. For instance, the max-plus polyhedron
$\max(x,y) \ge \max(-\infty,z)$ encodes the disjunctive information $(x < y
\wedge y \ge z) \vee (x \ge y \wedge x \ge z)$ or simply $y\ge z \;\vee \;
x \ge z$.  Max-plus polyhedra are the analogues of classical convex
polyhedra in the max-plus algebra, which operate over the reals and
$-\infty$ with $\max$ as the additive and $+$ as the multiplicative
operator~\cite{allamigeon2008inferring,kapur:arm13:geometric}.  
Dually, we also consider
min-plus polyhedra and 
combined max- and min-plus relations capturing 
if-and-only-if information.

We augment our dynamic analysis with a theorem prover based on
$k$-induction and SMT solving to verify candidate invariants.  Proven
results are true invariants of the program. 
Iterative reasoning using $k$-induction
allows us to prove invariants that cannot be proved using standard
induction, and in some cases to prove results that are not $k$-inductive.  
Moreover, recent advances in SMT solving
(e.g.,~\cite{jovanovic2012solving,calcs,demoura:tacas08:z3}) allow for efficient analysis over formulas in more
expressive logical theories, such as the theory of nonlinear arithmetic.

In summary, the paper makes the following contributions:

\begin{itemize}

\item A new algorithm to infer certain disjunctive
invariants dynamically by constructing nonconvex max- and min-plus polyhedra
over observed traces. 

\item The definition of a novel restricted class of max- and min-plus
invariants, called ``weak'' invariants, that strike a balance
between expressive power and computational complexity. Weak invariants 
express useful max- and min-plus relations and can be computed efficiently.

\item KIP, a theorem prover based on iterative
$k$-induction and SMT solving to verify dynamically inferred invariants
against program source code. When parallelized, KIP efficiently and correctly 
processes many complex and potentially spurious invariants.

\item An experimental evaluation on difficult kernels involving
nonlinear arithmetic and abstract arrays. Our approach is
efficient, both at learning disjunctive invariants and at proving them
correct. 
\end{itemize}

\iffalse
The rest of this paper is organized as follows: Section~\ref{motivation} shows how we generate disjunctive invariants from the execution traces of a simple program and verify the results.
Section~\ref{poly_invs} reviews how DIG uses different geometric shapes to analyze polynomial invariants and describes our extension to support disjunctive invariants of the max and min-plus forms. 
Section~\ref{k_ind} shows how KIP uses $k$-induction and SMT solving to verify candidate invariants and remove spurious results.
Section~\ref{results} reports experimental results. Section~\ref{related_work} surveys related work. 
Section~\ref{conclusion} concludes the paper.% and provides future directions.
\fi

\section{Motivating Example}\label{motivation}

We illustrate our methods with a simple example program containing a disjunctive invariant.

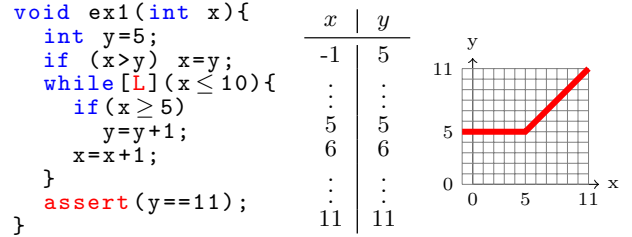
\begin{figure}[h]
\begin{minipage}{0.40\linewidth}
\begin{lstlisting}[numbers=none,mathescape,xleftmargin=0.0cm]
void ex1(int x){
  int y=5;
  if (x>y) x=y; 
  while[L](x$\,\le \,$10){
    if(x$\,\ge \,$5) 
      y=y+1;
    x=x+1;
  }
  assert(y==11);
}
\end{lstlisting}
\end{minipage}
\hspace{0.3cm} 
\begin{minipage}{0.0\linewidth}
\footnotesize
\centering
\begin{tabular}{c|c}
  $x$&$y$\\
  \midrule
-1 &5\\
$\vdots$&$\vdots$\\
5 &5\\
6 &6\\
$\vdots$ &$\vdots$\\
11 &11\\
\end{tabular}
\end{minipage}
\hfill
\begin{minipage}{0.38\linewidth}
\centering
\begin{tikzpicture}[scale=.14]
  \tikzstyle{every node}=[font=\tiny]
  \draw[->] (0,0) -- (11+1,0) node[anchor=west]  {\scriptsize x};
  \draw[->] (0,0) -- (0,11+1) node[anchor=south] {\scriptsize y};
  \draw[help lines] (-1,0) grid (11,11);
  \foreach \x in {0,5,11} \node[anchor=north] at (\x,0) {\scriptsize $\x$};
  \foreach \y in {0,5,11} \node[anchor=east]  at (-1,\y) {\scriptsize $\y$};
  \draw[red, line width = 0.08cm] (-1,5) -- (5,5) -- (11,11);
  % \draw[ultra thick] (5,5) -- (11,11);
  %\foreach \c in {(0,5),(5,5),(10,10)} \fill \c circle (5pt);
\end{tikzpicture}
\end{minipage}
\caption{Program \lt{ex1}, the observed traces on input $x=-1$, and the geometric representation of its invariant $(x < 5 \wedge y = 5) ~\vee~ (5 \le x \le 11 \wedge x = y)$ at location $L$.}
\label{ex1}
\end{figure}

Figure~\ref{ex1} shows program \lt{ex1}, adapted from
Gulwani and Jojic~\cite{Gulwani07programverification}.
The
program \lt{ex1} initializes $y$ to 5 and ensures $x \le y$, then enters a
loop that increments $y$ conditionally on the value of $x$.  Figure~\ref{ex1} also
shows the trace values for $x,y$ at location $L$ on the
input $x=-1$, and it depicts the nonconvex region
(a bent line) covering these trace points. 
Validating the postcondition \lt{y==11} requires analyzing
the semantics of the loop by identifying the invariants at $L$.

From the given trace, existing tools such as Daikon~\cite{daikon,ernstcollections} and
DIG~\cite{nguyen:icse12:dig,nguyen:tosem14:dig} can generate only conjunctive invariants such as 
\[
\small
\begin{array}{ccccc}
11 & \ge & x \\
11 & \ge & y & \ge & 5 \\
y  & \ge & x
\end{array} 
\] 
These relations are not expressive enough to capture
the disjunctive dependency between $x$ and $y$, and they fail to prove the
desired postcondition.

By building a max-plus polyhedra over the trace points in
Figure~\ref{ex1}, we obtain relations that simplify to: 
\[
\small
\begin{array}{ccccc}
11 & \ge & x & \ge & -1 \\
11 & \ge  & y & \ge & 5\\
0  & \ge  & x-y & \ge & -6\\
\end{array}
\]
\[
\small 
(x < 5 \wedge 5 \ge y) ~ \vee ~ (x \ge 5 \wedge x \ge y)
\] 
Note that the last relation is disjunctive. 
Next, we verify these candidate invariants against the
source code (Figure~\ref{ex1}) using $k$-induction and remove the spurious
relations $x\ge -1$ and $x-y \ge -6$. The rest are 
true invariants at $L$.

We note that the invariant $y \ge x$ is not directly $k$-inductive for
$k \le 5$.  However, by using the previously proven results 
$y \ge 5$ and $(x < 5 \wedge 5 \ge y) \vee (x \ge 5 \wedge x \ge y)$
as lemmas, 
our prover also verifies this relation $y \ge x$.  Further, 
the prover shows that $11 \ge x$ is redundant (i.e., implied by other
proved results) and can be removed.
The remaining invariants are:
\[
\small
\begin{array}{ccccc} 
11 & \ge & y   & \ge 5 \\
0  & \ge & x-y & \\ 
\end{array}
\]
\[
\small
(x < 5 \wedge 5 \ge y) ~ \vee ~ (x \ge 5 \wedge x \ge y)
\]
Intuitively, the code in Figure~\ref{ex1} has two phases: 
either $x<5$ (at which point the \lt{if} inside the \lt{while} loop is
not true and $y$ remains 5) 
or $x$ is between 5 and 11 
(at which point the \lt{if} inside the \lt{while} loop is
true, and $y=x$ because they are both incremented). The inferred
invariants are mathematically equivalent to the encoding of that intuitive explanation:  
\[
\small
\begin{array}{ccc} 
(x < 5 ~\wedge~ y = 5) & \vee & (5 \leq x \leq 11 ~\wedge~ y = x)
\end{array} 
\] 
They are also the precise
invariants of the loop and can prove the postcondition
\lt{y==11}. This example required that the dynamic analysis be expressive
and efficient enough to generate disjunctive invariants, and it
required that the 
static prover be expressive and efficient enough to remove spurious
invariants and prove the others correct. In the remainder of the
paper we describe these methods in detail.

\section{Invariant Inference Algorithm}\label{poly_invs}
This section describes our algorithm for inferring disjunctive
invariants from dynamic traces. We consider the construction of 
max-plus, weak max-plus, and min-plus invariants. 
We begin in Section~\ref{sec_dig} with a discussion of existing approaches
to inferring convex geometric invariants. 
We then explain how to apply such techniques to the inference of 
disjunctive invariants by reformulating the problem in the max-plus algebra
in Section~\ref{sec_mpi}.
In Section~\ref{mpInDig} we formalize that intuition and present our
algorithm pseudocode.
We then introduce in Section~\ref{weak_maxplus} an efficient and expressive restricted subclass, which we call weak max-plus invariants. 
In Section~\ref{odi} we describe the dual, min-plus invariants. 
Finally, in Section~\ref{analysis} we analyze the guaranteed properties of
our algorithm.

\subsection{Inferring Convex Geometric Invariants}\label{sec_dig}

\begin{figure}[h]
\centering
\begin{minipage}{1.0in}
  \centering
  \includegraphics[width=0.72\linewidth]{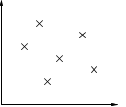}
  %\caption{a}
\end{minipage}
\begin{minipage}{1.0in}
  \centering
  \includegraphics[width=0.72\linewidth]{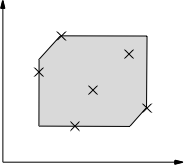}
  %\caption*{b}
\end{minipage}
\begin{minipage}{1.0in}
  \centering
  \includegraphics[width=0.72\linewidth]{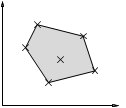}
  %\caption*{c}
\end{minipage}
\caption{(a) A set of points in 2D and its approximation using a (b) zone
and (c) polygon region. 
}
\label{geo-shapes}
\end{figure}

We review the problem of learning convex
geometric invariants. Figure~\ref{geo-shapes} visualizes these invariants,
showing 2D points (panel a) and two examples of increasingly 
precise, but also increasingly expensive, convex shapes containing 
those points (panels b and c).  The inferred invariants correspond to
the relations defining the enclosing shapes given the trace points
from panel a.

The dynamic analysis tool DIG~\cite{nguyen:icse12:dig,nguyen:tosem14:dig} 
generates different forms of polynomial invariants by
building geometric shapes, such as those shown in
Figure~\ref{geo-shapes}, enclosing the trace points.  
DIG first determines if the points lie in a simple hyperplane.
If such a plane does not exist, DIG then computes a
convex hull over the trace points.  Such hulls are bounded convex
polyhedra---each polyhedron is enclosed by a finite number of facets
and contains the line joining any pair of its points.  The
\emph{half-space} representation of such a polyhedron is a set of finite
linear relations of the form $c_1v_1 +\dots + c_nv_n \ge c_0$.  The facets
of the polyhedron, corresponding to the solutions of the set of linear
inequalities, give a set of candidate inequality invariants among the variables
$v_i$.  Figure~\ref{geo-shapes}c shows a 2D polyhedron with five facets,
represented by five linear inequalities. For efficiency, DIG also considers
more restricted forms of inequalities representing simpler geometric
shapes, such as the six-edged zone relation~\cite{CCF05,mine2004weakly}
in Figure~\ref{geo-shapes}b and the eight-edged octagon relation~\cite{CCF05}.

To support nonlinear relations, DIG lifts its analysis to terms 
representing nonlinear polynomials over program variables. For example,
rather than analyzing variables $v_1$ and $v_2$ directly, 
relations can be constructed among the terms $t_1 = v_1, t_2 = v_1v_2$ (note
that $t_2$ is nonlinear). Thus, equations such as $t_1 + t_2=1$ can
be generated, which represents a
line over $t_1,t_2$ but a hyperbola over $v_1,v_2$.  When additional traces
are available, filtering step removes spurious invariants~\cite{nguyen:icse12:dig}.

These existing approaches lack the expressive power to
learn disjunctive invariants, a gap that we address 
in the following subsection by reformulating in the
max-plus algebra.

\subsection{Max-Plus Invariants}\label{sec_mpi}

As discussed earlier, programs containing loops or conditional branches are not adequately modeled by purely conjunctive invariants.
Figure~\ref{ex1} depicts the nonconvex region defined by the loop invariant  $(x < 5 \wedge y = 5)  \vee (5 \le x \le 11 \wedge x = y)$ in our program \lt{ex1}.
Such disjunctive information cannot be expressed as a conjunction of polynomial relations, including octagon or even general polyhedron forms.
Although disjunctive invariants can be simulated using polynomials of higher order (e.g., $a = 0 \vee b = 0$ is equivalent to $a \times b = 0$), this approach generates terms with impractically high degree and  computational cost, especially when there are more than two disjunctions.
We thus require a fundamentally different approach.

To model disjunctive invariants, we use formulas representing \emph{max-plus} polyhedra~\cite{allamigeon2008inferring}, i.e., nonconvex hulls that are convex over a max-plus algebra. 
Max-plus formulas allow disjunctions of zone relations~\cite{CCF05,mine2004weakly}: inequalities of the forms $\pm v \ge c$ and $v_1 - v_2 \ge c$.
Formally, max-plus relations have the structure 
\[
\begin{array}{l}
\max(c_0,c_1+v_1,\dots,c_n+v_n) \ge \\
\max(d_0,d_1+v_1,\dots,d_n+v_n),
\end{array}
\]
where $v_i$ are program variables, $c_i,d_i$ are real numbers or $-\infty$, and $\max(t_0,\dots,t_m)$ returns the largest $t_i$. That is, $\max(x,y) \equiv \mathsf{if}\; x > y \;\mathsf{then}\; x\; \mathsf{else}\; y$.
We note that $\max(v_0, v_1-\infty,v_2,\dots, v_n ) = \max(v_0,v_2, \dots, v_n)$ and thus we often drop $-\infty$ $\max$-arguments.

The $\max$ operator allows max-plus formulas to encode certain disjunctions.
For example, the max-plus relation $\max(0,x-5,y-\infty) = \max(-\infty,x-\infty,y-5)$, i.e., $\max(0,x-5) = \max(y-5)$, encodes the disjunction $(5 > x  \wedge y=5) \vee (5 \le x \wedge x=y)$, or $y=5 \vee x=y$.\footnote{For presentation purpose we abbreviate max-plus notations, e.g.,  $\max(x,y)\ge z$ for $\max(x,y,z-\infty,-\infty) \ge \max(x -\infty,y-\infty,z,-\infty)$ and $x\ge 9$ for $\max(9,x-\infty,y-\infty) \ge \max(-\infty,x,y-\infty)$. An equality is also used to express the conjunction of two inequalities, e.g.\  $\max(x,y)=z$ for $\max(x,y)\ge z \wedge z \ge \max(x,y)$.}

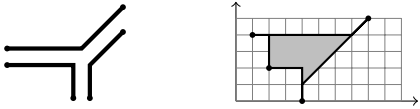
\begin{figure}[h]
\centering
\begin{minipage}{0.40\linewidth}
\centering
\begin{tikzpicture}[scale=.22]
  \tikzstyle{every node}=[font=\small]
  \foreach \c in {
    (0,3),  (7,5.5),   %top
    (0,2),  (4,0),     %left
    (5,0),  (7,4)      %right
  } \fill \c circle (5pt);

  \draw[ultra thick] (0,3) -- (4.5,3) -- (7,5.5); %top
  \draw[ultra thick] (0,2) -- (4,2) -- (4,0); %left
  \draw[ultra thick] (5,0) -- (5,2) -- (7,4); %right
\end{tikzpicture}
%\caption*{a}
\end{minipage}
\begin{minipage}{0.40\linewidth}
\centering
\begin{tikzpicture}[scale=.22]
  \tikzstyle{every node}=[font=\small]
  \draw[->] (0,0) -- (10+1,0);
  \draw[->] (0,0) -- (0,5+1);
  \draw[help lines] (0,0) grid (10,5);
  \draw[thick,fill=black!25] (1,4) -- (7,4) -- (8,5) -- (4,1) -- (4,0) -- (4,2) -- (2,2) -- (2,4);
  \foreach \c in {(1,4),(2,2),(4,0),(8,5)} \fill \c circle (5pt);
\end{tikzpicture}
%\caption*{b}
\end{minipage}
\caption{(a) Three possible shapes of a max-plus line segment: $\max(x+a,b) \ge y$ (top), $\max(y+a,b) \ge x$ (right), $\max(x+a,y+b) \ge c$ (left) and (b) a max-plus convex hull built over four points using these line segments.}  %VU: todo, are these formulas correct ?
\label{fig_maxp}
\end{figure}

Max-plus relations are analogous to polyhedra relations, but use $(\max,+)$ instead of the $(+, \times)$ of standard arithmetic.
These operators allow max-plus relations to form geometric shapes that are nonconvex in the classical sense. 
For example, the max-plus relation $x=y\vee y=5$ represents a nonconvex region consisting of two lines $x=y$ and $y=5$.  
Moreover, the structure of max-plus relations produces a relatively peculiar geometric shape.  
Figure~\ref{fig_maxp}a shows the three possible shapes of a max-plus line segment in 2D.  
In general dimensions, two points are always connected by lines that run parallel,
perpendicular, or at a 45 degree angle to all the coordinate axes.  A
max-plus polyhedron consists of these connections and the area surrounded
by them. Figure~\ref{fig_maxp}b depicts a max-plus polyhedron represented
by a set of four lines connecting the four marked points.  Although a
max-plus polyhedron is not convex in the classical sense, it is convex in
the max-plus sense using max-plus algebra.  That is, it contains
any max-plus line segment between any pair of its points.  This allows us
to generate max-plus polyhedra over a finite set of
traces, as shown in Figure~\ref{fig_maxp}b. Where there is no confusion,
we shorten max-plus (resp. min-plus) to max (resp. min) when describing
polyhedra, formula, or relations. 

A bounded max polyhedron can have finitely many facets representing max relations, e.g., even a 2D complex polygon may contain multiple edges.
Thus, a disjunctive formula representing a max polyhedron has no fixed bounds on the number of disjuncts used.
However, computing a max polyhedron over $n$ points in $d$
dimensions is computationally expensive
$O(n^d)$~\cite{allamigeon2008inferring}, similar to classical
polyhedron computations. 
Next, we propose heuristics to avoid
generating these high-dimensional polyhedra in Section~\ref{mpInDig}.
Section~\ref{weak_maxplus} then introduces a simpler form of max relations
that strikes a reasonable compromise between efficiency and precision.

\subsection{Dynamically Inferring Max-Plus Invariants}\label{mpInDig}

\begin{algorithm}
\small
\DontPrintSemicolon
\SetKwInOut{Input}{input}
\SetKwInOut{Output}{output}
\SetKwFunction{genTerms}{genTerms}
\SetKwFunction{extractFacets}{extractFacets}
\SetKwFunction{genPoints}{genPoints}
\SetKwFunction{createPolyhedron}{createMaxPlusPolyhedron}
\SetKwFunction{deduceIeqs}{deduce$_{\text{ieqs}}$}
\SetKwFunction{genInvsEqts}{genInvs$_{\text{eqts}}$}

\SetKwData{None}{None}
\SetKwData{ieqs}{ieqs}
\SetKwData{eqts}{eqts}
\SetKwData{True}{True}

\Input{set of variables $V$, set of traces $X$, max degree $d$}
\Output{set $S$ of polynomial inequalities}
\BlankLine

$T \leftarrow$ \genTerms{$V$,$d$}\;
$P \leftarrow$ \genPoints{$T$,$X$}\;
$H \leftarrow$ \createPolyhedron{$P$}\;
$S \leftarrow$ \extractFacets{$H$}\;
\KwRet{$S$}
\caption{High-level algorithm for finding disjunctive polynomial inequalities.}
\label{proc-mp}
\end{algorithm}

We infer max invariants dynamically using a procedure similar to that
used for classical polyhedra invariants. 
Figure~\ref{proc-mp} outlines the main steps of the algorithm: using terms
to represent program variables; instantiating points
from terms using input traces; creating a 
max polyhedron enclosing the points;
and extracting its facets to represent max relations among terms.

Because program invariants often involve only a small
subset of all possible program variables, we employ heuristics to
search iteratively for invariants containing all possible combinations of
a small, fixed number of variables. We propose to consider
max relations over triples of
program variables, i.e., 
$\max(c_0,c_1+v_1,c_2+v_2,c_3+v_3)\ge \max(d_0,d_1+v_1+d_2+v_2,d_3+v_3)$
representing max polyhedra in three-dimensional space. %VUCOMMENT913: 3 vars - 3 dims

Our algorithm also supports nonlinear max relations by using terms to represent nonlinear polynomials over variables.
However, the number of possible terms is exponential in the number of
degrees~\cite{nguyen:icse12:dig} and thus we target linear max relations by default for
efficiency.
The user of DIG can change the parameter $d$ in Figure~\ref{proc-mp} to generate
higher degree relations (e.g., $d=2$ for quadratic relations) and can also
manually define terms to capture other desirable properties.
For example, a user with knowledge about the shape of the
desired invariants might hypothesize a \emph{spherical} shape
$\max(c_0,c_1+x^2,c_2+y^2)\ge \max(d_0,d_1+x^2,d_2+y^2)$.  With that
as input, the algorithm searches for that exact shape (i.e., computes
the coefficients $c_i,d_i$) from the polyhedron built over the trace
points of the terms representing the nonlinear polynomials $x^2$ and
$y^2$.

\subsubsection*{Example} We illustrate the algorithm by deriving the
invariant $(x < 5 \wedge y = 5)  \vee (5 \leq x \leq 10 \wedge x = y)$
at location $L$ in program \lt{ex1} in Figure~\ref{ex1}.
The trace values for $x,y$ in Figure~\ref{ex1} 
form a set of eleven points (e.g., the first is $(-1,5)$).
We then compute a max polyhedron 
over these points.  The half-space representation of that polyhedron
consists of the max relations:
\[
\small
\begin{array}{ccccc} 
11 & \ge & x & \ge & -1 \\
11 & \ge & y & \ge & 5 \\ 
0  & \ge & x-y & \ge & -6 \\
\max(0, x-5) & \ge & y - 5 
\end{array} 
\] 
The conjunction $0 \ge x-y \;\wedge\; 11\ge x\; \wedge\; \max(x-5,0)
\ge y - 5$, which forms the nonconvex region in Figure~\ref{ex1}, is
logically equivalent to the invariant $(x < 5 \wedge 5 = y)  \vee (5
\leq x \leq 11 \wedge x = y)$.

Note that  $x\ge -1$ and $x-y \ge -6$ are spurious relations because $x$
has no lower bound.  Additional traces, such as running \lt{ex1} on
$x=-5$, would remove these spurious invariants.  More generally, the static
technique in Section~\ref{k_ind} formally verifies candidate invariants and
removes spurious results.

\subsection{Weak Max-Plus Invariants}\label{weak_maxplus}

We introduce and define a weaker form of max relations that
retains much expressive power but avoids the high computational cost
of computing a max polyhedron.
Our approach is inspired by earlier methods for finding simpler forms of
inequalities (e.g., zone and octagon) to avoid the cost of
finding general polyhedra~\cite{mine2004weakly}.
To the best of our knowledge, this is the first attempt to consider a
simpler form of max inequalities for program analysis.

We define a \emph{weak} max relation to be of the form: 
\[
\begin{array}{l}
\max(c_0,c_1+v_1,\dots,c_k+v_k) \ge v_j+d,\\ 
v_j + d \ge \max(c_0,c_1+v_1,\dots,c_k+v_k),
\end{array}
\]
where $v_i$ are program variables, $c_i \in \{0,-\infty\}$, $d$ is a real
numbers or $-\infty$, and $k$ is a constant, e.g., $k=2$. Unlike general
max relations, weak max relations have some convenient properties:
\begin{enumerate}
  \item They restrict the values of the coefficients $c_i$ to $\{0,-\infty\}$.
    The general form allows $c_i \in \mathbb{R} \cup \{-\infty\}$. 
  \item They fix the number of variables $k$ to a small constant. The general
  form allows $n$ variables. 
  \item They allow only one unknown parameter $d$. The general form
  allows $d_0 \dots d_n$. 
\end{enumerate} 
Weak max relations are thus a strict subset of general max
relations. For example, the weak max form cannot represent 
general max relations like $\max(x+7,y)\ge z$ or $\max(x,y) \ge
\max(z,w)$, but it does support zone relations like $x -y \ge 10, x = y$
and disjunctive relations like $\max(x,y)\ge z$ and $\max(x,0) \ge y+7$.

Geometrically, weak max relations are a restricted kind of
max polyhedra.
While general max line segments have the possible three
shapes shown in Figure~\ref{fig_maxp},  weak max line
segments have only two shapes represented by the formulas $\max(x,b) \ge y$
and $\max(y,b) \ge x$. That is, weak max shapes include only lines
that run in parallel or at a 45 degree
angle. Lines with a perpendicular shape cannot occur because their formula,
$\max(x,y)\ge 0$, is inexpressible using the weak max form.  

The advantage of these restrictions is that they admit a
straightforward algorithm to compute the bounded weak max 
polyhedron over a set of finite points in $k$ dimensions. The
algorithm first enumerates all possible weak relations over $k$
variables and then finds the unknown parameter $d$ in each relation
from the given points.  The resulting set of relations is the
half-space representation of the weak max polyhedron enclosing the
points.

Note that this algorithm does not apply to the general max form because the
coefficients $c_i$ are not enumerable over the reals.  Moreover, the problem
becomes more complex when more than one unknown is involved.  For instance, it
is nontrivial to compute the unknowns $c,d$ in the max relation
$\max(c,x) \ge y + d$ because the values of $c$ and $d$ depend on each other.

\subsubsection*{Example} 
We illustrate this algorithm by finding the weak max 
polyhedron
enclosing the points $\{(x_1,y_1), \dots, (x_n,y_n)\}$ in 2D.
First, we enumerate relations of the weak max form by instantiating the coefficients
$c_i$ over
$\{0,-\infty\}$.  For the form $\max(c_0,c_1+x,c_2+y) \ge x+d$ we obtain
eight max relations (two choices each for three coefficients): 
\[
\small
\begin{array}{lr} 
\max(0,x,y)\ge x+d, &\dots\\
\max(0,x) \ge x+d,  & -\infty \ge x+d \\
\end{array} 
\]
The eight additional max relations for each of the other three forms
$\max(c_0,c_1+x,c_2+y) \ge y+d$, $x + d\ge \max(c_0,c_1+x,c_2+y)$, $y+d \ge
\max(c_0,c_1+x,c_2+y)$ are obtained similarly.  
Redundant relations can be removed (e.g., $\max(y,0)\ge x$
implies $\max(x,y,0)\ge x$).  

Next, we compute the parameter $d$ in each of the 32 obtained relations
using the given points $\{(x_1,y_1), \dots, (x_n,y_n)\}$.  For instance,
$\max(y,0)\ge x + d$ has $d = \min(\max(y_i,0)-x_i)$ and  $x +d \ge
\max(y,0)$ has $d=\max(\max(y_i,0)-x_i)$.  The resulting relations form an
intersecting region that represents a bounded weak max polygon over
the given points.

In general, the number of  weak max relations enumerated over $k$
variables is $O(k 2^{k+2})$ and the time to find the 
single parameter $d$ in each relation is linear in the number of points.
Thus, the complexity for computing a weak max polyhedron over $n$ points
in $k$ dimensions is $O(n 2^k)$. The complexity is therefore 
polynomial in the number of points when $k$
is a constant and is exponential in the number of dimensions when $k$ is not
fixed. Note that even this worst case is still smaller than $O(n^d)$, the
complexity of building a general max polyhedron.
Importantly, the number of facets of a weak max polyhedron has a fixed 
upper bound for
each $k$. For example, $k=2$ has at most 32 facets. 
This is thus more manageable than the number of facets of
general max polyhedron, which can be arbitrarily finitely many.

\subsection{Min-Plus Invariants}\label{odi}

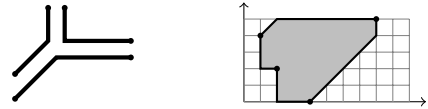
\begin{figure}[h]
\centering
\begin{minipage}{0.40\linewidth}
\centering
\begin{tikzpicture}[scale=.22]
  \tikzstyle{every node}=[font=\small]
  \foreach \c in {
    (3,5.5),  (7,3.5),   %top
    (2,5.5),  (0,1.5),     %left
    (0,0),  (7,2.5)      %right
   } \fill \c circle (5pt);

   \draw[ultra thick] (3,5.5) -- (3,3.5) -- (7,3.5); %top
   \draw[ultra thick] (2,5.5) -- (2,3.5) -- (0,1.5); %left
   \draw[ultra thick] (0,0) -- (2.5,2.5) -- (7,2.5); %right
\end{tikzpicture}
%\caption*{a}
\end{minipage}
\begin{minipage}{0.40\linewidth}
\centering
\begin{tikzpicture}[scale=.22]
  \tikzstyle{every node}=[font=\small]
  \draw[->] (0,0) -- (10+1,0);
  \draw[->] (0,0) -- (0,5+1);
  \draw[help lines] (0,0) grid (10,5);
  %\draw[thick,fill=black!25] (1,4) -- (7,4) -- (8,5) -- (4,1) -- (4,0) -- (4,2) -- (2,2) -- (2,4);
  \draw[thick,fill=black!25] (1,4) -- (2,5) -- (8,5) -- (8,4) -- (4,0) -- (2,0) -- (2,2) -- (1,2) -- (1,4);
  \foreach \c in {(1,4),(2,2),(4,0),(8,5)} \fill \c circle (5pt);
\end{tikzpicture}
%\caption*{b}
\end{minipage}
\caption{(a) Three possible shapes of a min-plus line segment and (b) a min-plus polyhedron built over four points.}
\label{fig_minp}
\end{figure}

We also consider \emph{min} relations of the form  
\[
\begin{array}{l}
\min(c_0,c_1+v_1,\dots,c_n+v_n) \ge \\
\min(d_0,d_1+v_1,\dots,d_n+v_n), 
\end{array}
\]
where $v_i$ are program variables and $c_i,d_i \in \mathbb{R} \cup
\{\infty\}$.
Similar to its max dual, a min polyhedron is a 
formed by the intersection of finite min lines.
However, min and max relations describe different forms of disjunction
information and have different geometric shapes.
For instance, the relation $\min(x,y)=z$ encodes the disjunction $(x < y \Rightarrow x = z) \wedge (x\ge y \Rightarrow y = z)$ that is not expressible 
as a max relation.
Figure~\ref{fig_minp} depicts the min version of the shapes in
Figure~\ref{fig_maxp}.

A conjunction of max and min invariants can describe 
information that is inexpressible using either max or min relations
alone. Consider  program \lt{ex2} in Figure~\ref{ex2}, which has the
invariant $y\le 10 \Leftrightarrow b=0$ at location $L$. 
By building max and min polyhedra over the
traces
given in Figure~\ref{ex2}, we obtain $1 \ge b \ge 0$, $\max(y - 10,0) \ge
b$, and $b + 10 \ge \min(y,11)$.
Given $1 \ge b \ge 0$, the max relation implies $
b = 0 \Rightarrow y \le 10$ and 
the min
relation implies $b \neq 0 \Rightarrow y > 10$.
These disjunctions are mathematically equivalent to the $\mathsf{iff}$ condition
$y\le 10 \Leftrightarrow b=0$.

\begin{figure}[h]
\begin{minipage}{0.45\linewidth}
\begin{lstlisting}[numbers=none,mathescape,xleftmargin=1.25cm]
int ex2(int x){ 
  int y, b; 
  if (x>=0) {y=x+1;}
  else {y=x-1;}
  b=(y>10);
  [L]
  return b;
}
\end{lstlisting}
\end{minipage}
\hfill
\begin{minipage}{0.45\linewidth}
\small
\centering
\begin{tabular}{c|c c}
  $x$&$y$&b\\
  \midrule
  -50&-51&0\\
  -33&-34&0\\
   9&10&0\\
  10& 11& 1\\
  12& 13& 1\\
  40&41&1\\
\end{tabular}
\end{minipage}
\caption{Program \lt{ex2} and its trace data at location $L$ for several input values.}
\label{ex2}
\end{figure}

Dually, we also define weak min relations:
\[
\begin{array}{l}
\min(c_0,c_1+v_1,\dots,c_k+v_k) \ge v_j+d_i,\\
v_j + d_i \ge \min(c_0,c_1+v_1,\dots,c_k+v_k),
\end{array}
\]
where $v_i$ are program variables,  $k$ is a constant, $c_i \in \{0,-\infty\}$, and $d_i \in \mathbb{R} \cup \{-\infty\}$.
The algorithm for computing weak min
polyhedra over finite points is similar to the one for weak max polyhedra and has
equivalent theoretical complexity as given in Section~\ref{weak_maxplus}.

\subsection{Algorithmic Analysis}\label{analysis}

We analyze important properties associated with our
algorithm. There are two key concerns: the production of
spurious invariants that underapproximate general program behavior
(too strong relations that hold only for some inputs)
and invariants that overapproximate\footnote{For instance,
if the true program behavior is $x \le y - 5$, the weaker candidate
invariant $x \le y$ is a strict overapproximation: it is always true
($x \le y - 5 \Rightarrow x \le y - 0$) but is not the most precise
answer.} 
general program behavior
(weak invariants that may not be useful). 

By building max or min polyhedra over trace points, 
which are convex in the corresponding max or min algebra,
our algorithm guarantees that
it produces candidate invariants that always underapproximate, or are
equivalent to, program invariants expressible under the max or min-plus
forms. The proof of this claim follows from the facts that (1) the given
set of observed traces is a subset of all possible program traces and (2)
our constructed max polyhedron over a set of points is the smallest 
 max polyhedron represented by those points. The proof details 
follow those of the underapproximation argument for inferring
classically convex shapes~\cite{nguyen:tosem14:dig}. Thus, if the program invariants 
are expressible in our system, our algorithm never overapproximates. 

This underapproximation property is important because its violation---a
candidate invariant strictly overapproximating the program
invariant---indicates a bug in the subject program. For example, if
the expected invariant is $t\ge 1$ but our algorithm discovers $t \ge
0$, then the underapproximation property guarantees the value $t = 0$
exists in the observed traces. The trace with $t=0$ represents a
counterexample that violates the expected property of $t$ being
positive.

Underapproximation properties also represent spurious invariants.
One way of understanding why spurious results occur is that 
(1) a max or min polyhedron has many facets in
high-dimensional space, and (2) inadequate traces may result in 
a constructed polyhedra with facets
representing spurious relations.  For instance, if $x,y$ can take any value
over the reals, then an $n$-facet max polygon built over
any set of trace points for $x,y$ produces $n$ spurious invariants because
no bounded max-plus polygons can capture the unbounded ranges of $x,y$. 
Both filtering against additional traces and restricting attention to 
the weaker forms of max and min relations help reduce 
spurious invariants. In the next section we describe a more
general technique, based on theorem proving, to distinguish
between true and spurious invariants.

\section{Verifying Candidate Invariants}\label{k_ind}

Our algorithm, and convex hull methods in general, can generate many
powerful but potentially incorrect relations due to trace incompleteness.
We augment dynamic invariant generation with static theorem proving to
produce sound program invariants with respect to the program source
code.
Specifically, we verify program invariants using $k$-induction. In this
approach, 
$k$ base cases are specified, and the $k$ previous instances are
available for
proving the inductive step  (e.g.,~\cite{donaldson11}).  This
additional
power allows us to prove many invariants relevant
to program verification that do not admit standard induction.

Our theorem prover design, called KIP, is based on iterative $k$-induction
and uses SMT solving to verify candidate invariants. 
In addition, its architecture supports parallel
checking of invariants, dramatically improving efficiency. Recent advances
in SMT solving~\cite{jovanovic2012solving,calcs,demoura:tacas08:z3} allow for efficient analysis over formulas
encoding complex programs and properties in powerful theories. This means
that we can reason about, and verify, invariants involving theories
such as nonlinear arithmetic and data structures such as arrays, bit
vectors, and pointers.

\begin{wrapfigure}{r}{0.20\textwidth}
\vspace{-4ex}
\centering
\begin{lstlisting}[numbers=none,escapechar=!,xleftmargin=-0.20cm]
int sqrt(int x){
  assert(x>=0);
  int a=0,s=1,t=1;
  while[L](s<=x){
    a += 1;
    t += 2;
    s += t;
  }
  return a;
}
\end{lstlisting}
\vspace{-4ex}
\end{wrapfigure}
Consider the program \lt{sqrt} on the right, which computes the square root
of an integer using only addition. 
From observed traces at location $L$, our algorithm generates candidate
loop invariants such as $t =  2a + 1, 4s = t^2 + 2t + 1, s = (a + 1)^2,
s\ge t$ and $x \le 9989$.
KIP successfully distinguishes true and false invariants from these
results. Specifically, we prove $t=2a+1$ and $4s = t^2 + 2t + 1$ are
inductive invariants and $s = (a+1)^2$ is a $1$-inductive invariant (i.e., cannot be proved using standard induction).
By using proved results as lemmas, KIP is able to show
the invariant $s \ge t$, which is not $k$-inductive for $k \le \mathsf{maxK}$, where $\mathsf{maxK}=5$ is the default setting of KIP.
The prover also rejects spurious relations such as $x\le 9989$ by
producing counterexamples that invalidate those relations in \lt{sqrt}.
The parallel implementation allows the prover to check these candidate results 
simultaneously.

\subsection{Analyzing Programs using  $k$-Induction}

A program execution can be modeled as a state transition
system $M=(I,T)$ with $I$ representing the initial state of $M$, and
$T$
specifying the transition relation of $M$ from a state $n-1$ to a state $n$.
To prove that $p$ is a \emph{state invariant} that
holds at every state of $M$, $k$-induction requires that $p$ hold for the
first $k+1$ states (base case) and that $p$ hold for the state $n+k+1$
assuming that it holds for the $k+1$ previous states (induction step).
Formally, $k$-induction proves the state invariant $p$ of
$M=(I,T)$ by checking the base case and
induction step formulas:

\footnotesize
\begin{align}
  I \wedge T_1 \wedge \dots \wedge T_k \;&\Rightarrow \; p_0 \wedge \dots \wedge p_k\label{kind-1}\\
  p_n \wedge T_{n+1} \wedge  \dots \wedge p_{n+k} \wedge T_{n+k+1} \;&\Rightarrow \; p_{n+k+1}\label{kind-2}
\end{align}
\normalsize

\noindent If both formulas hold then $p$ is a $k$-inductive invariant. 
If the base case~\eqref{kind-1} fails then $p$ is disproved and thus is not
an invariant (assuming that $M$ correctly models the program).
However, if the base case holds but the induction step~\eqref{kind-2}
fails, then $p$ is not a $k$-inductive invariant, but it could still
be a program invariant.
Thus, $k$-induction is a sound but incomplete proof technique.

By considering multiple consecutive transitions, $k$-induction can prove
invariants that cannot be proved by standard induction ($0$-induction
in this formulation). 
For instance, the invariant $x\ne y$ of the machine $M(I: (x=0 \wedge
y=1 \wedge z=2)_0$, $T_n: x_n = y_{n-1} \wedge y_n = z_{n-1} \wedge z_n =
x_{n-1}$) that rotates the values $0,1,2$ through the variables $x,y,z$ is
not provable by
standard induction but is $k$-inductive with $k \ge 3$.  
The notation $(P)_i$ denotes the formula $P$ with all free variables
subscripted by $i$, e.g., $(x + y = 1)_0$ is $x_0 + y_0 = 1$.

\subsection{$k$-Induction and SMT Solving}

\begin{algorithm}[h!]
\small
\DontPrintSemicolon
\SetKwInOut{Input}{input}
\SetKwInOut{Output}{output}

\Input{$\mathit{I, T, p}$}
\Output{$\{\mathit{proved,disproved,unproved}\}$}
\BlankLine

\BlankLine
\For{$\mathit{k=0}$ \KwTo $\mathsf{maxK}$}{

  \tcp{base case}

  \lIf{$\mathit{k=0}$}{$\mathit{S_b.assert(I)}$} \lElse{$\mathit{S_b.assert(T_k)}$}

  \lIf{$\mathit{\neg S_b.entail(p_k)}$}{
    \KwRet ($\mathit{disproved, S_b.cex}$)
  }

  \BlankLine
  \tcp{induction step}
  $\mathit{S_s.assert(p_k,T_{k+1})}$\;
  \lIf{$\mathit{\neg S_s.entail(p_{k+1})}$}{\Return $\mathit{proved}$}
}

\BlankLine
\KwRet $\mathit{unproved}$% \;
\caption{Procedure $\mathsf{kprove}$ for incremental $k$-induction using SMT solvers
$S_b$ and $S_s$.}
\label{kind-base}
\end{algorithm}

Figure~\ref{kind-base} outlines the procedure for verifying a property $p$ using inductive $k$-induction with SMT solving.
The procedure consists of a loop that performs incremental $k$-induction, starting from $k=0$.
The loop
terminates when either the base case fails ($P$ is not an invariant), both
the base case and the induction step hold ($P$ is an invariant), or
$\mathsf{maxK}$ is reached ($P$ is not a $\mathsf{maxK}$-inductive
invariant).

 %In the first case, $P$ is not an invariant and n the last case, we fail to decide $P$ within the given $K$ and therefore the status of $P$ is unknown.  

We use two independent SMT solvers $\mathit{S_b}$ and $\mathit{S_s}$
to check the two formulas corresponding to the base case~\eqref{kind-1} and
induction step~\eqref{kind-2}.\footnote{The two SMT solvers can share the
same implementation: ``independent'' merely indicates that they may
hold different assumptions at runtime.} 
For a solver $S$ and a formula $f$, we append $f$ to $S$ through
\emph{assertion} and check if the assertions in $S$ imply $f$ using
\emph{entailment}~\cite{demoura:tacas08:z3}. 
If $S$
does not entail $f$, then the solver returns a counterexample (cex)
satisfying $a_1\wedge \dots\wedge a_n$ but not $f$.

\iffalse
Counterexamples are useful for debugging because a cex represents a set of
variable values that invalidates a formula.  Specifically, the cex
$\mathit{S_b.cex}$ represents a concrete execution sequence that shows the
failure of $P$ within $k$ transitions.  Moreover, considering $k$ in
ascending order means that $\mathit{S_b.cex}$ has the fewest number of
transitions explicating the violation of $P$ from the initial state of the
machine.
\fi

\subsection{The Architecture of KIP}\label{verify_with_KIP}
\begin{algorithm}
\small
\DontPrintSemicolon
\SetKwInOut{Input}{input}
\SetKwInOut{Output}{output}

\Input{$\mathit{S,L,P}$}
\Output{$\mathit{P_i, P_r, P_d, P_u}$}
\BlankLine

$\mathit{I,T \leftarrow vcgen(S,L)}$\;

$\mathit{P_p \leftarrow \emptyset; P_d \leftarrow  \emptyset; P_u \leftarrow \emptyset}$\;

\Repeat{$\mathit{New_p= \emptyset \;\vee\; New_u = \emptyset}$}{

    $\mathit{New_p \leftarrow \emptyset; New_u \leftarrow \emptyset}$\;

   \ForEach{$\mathit{p \in P}$}{
     $\mathit{r \leftarrow kprove(I,T,p)}$\;

    \If{$\mathit{r = proved}$}{
      $\mathit{P_p.add(p)}; \mathit{New_p.add(p)}$\;
     }
     \lElseIf{$\mathit{r = unproved}$}{
       $\mathit{New_u.add(p)}$\;
     }
    \lElse{$P_d.add(p)$\;}
    
    }

   $\mathit{KIP.addLemmas(New_p)}$\;
   $\mathit{P \leftarrow New_u}$\;
}
$\mathit{P_u \leftarrow P}$\;
$P_i,P_r = check\_redundancy(P_p)$\;
\KwRet{$P_i, P_r, P_d, P_u$}\;
\caption{Procedure to verify candidate invariants. $P_i$ and $P_r$ are proved results, however $P_r$ are redundant because $P_i \Rightarrow P_r$. $P_d$ and $P_u$ are disproved and unknown results, respectively.}
\label{proc_verify}
\end{algorithm}

At a high level, verifying a candidate invariant against a program requires
two steps: (1) computing a formula that encodes the program's semantics; and (2) proving whether the candidate invariant is consistent
with that formula or not. To increase expressive power in practice, we also
(3) incorporate knowledge of all invariants learned thus far. 

Figure~\ref{proc_verify} outlines the architecture of KIP, our $k$-inductive parallel theorem prover, to verify a set $P$ of candidate obtained at location $L$ for program $S$.
We first generate from the program $S$ and the location $L$ the formulas
$I,T$.  These formulas can be thought of as representing the state
transition system $M=(I,T)$ described above. 
Equivalently, $I, T$ can be thought of as
verification conditions (vcs) based on 
weakest preconditions (wps) from program analysis using Hoare logic.

The backward analysis method~\cite{dijkstra:cacm75:wp} provides the necessary rules
to create $I,T$ for imperative programming constructs such as assignments,
conditional branches, and loops. This area is well-established---tools such
as Microsoft Boogie~\cite{leino2008boogie} and ESC~\cite{detlefs1998extended} implement various
methods based on backward analysis to automatically generate vcs using wps.

Our algorithm progresses by trying to prove the invariants in the context
of the vcs. While unproved invariants remain, the procedure
attempts to re-prove them by adding newly proved results as lemmas to KIP. 
In many cases, this additional knowledge allows KIP to prove properties
that could not be proved previously (see Sections~\ref{motivation} and
\ref{sec-rq2}).
A disproved invariant is likely spurious (e.g.,
assuming $I,T$ correctly models the program), a proved invariant is
definitely correct, and an unproved invariant (e.g., one that is not
$\mathsf{maxK}$-inductive) can be conservatively rejected.

The algorithm supports parallelism, which can check candidate invariants (the {\tt for} loop in Figure~\ref{proc_verify}) simultaneously using multiple threads.
In a post-processing step, KIP uses implication to partition
all proved invariants into two sets: those that are independent (i.e.,
strongest) and those that can be implied by the others (i.e., weaker). 
The implied invariants are redundant and need not be presented to
the developer. This partitioning uses the backend SMT solver to check if each invariant $p\in P_p$ can be inferred by the conjunction of other proved invariants $P_p \setminus \{p\}$.

Overall, KIP's design represents a novel combination of established
techniques and provides the five
properties we desire for the efficient verification of complex
invariants: (1) use of $k$-induction for expressive power; (2) use of SMT solvers for reasoning about program-critical theories like
nonlinear arithmetic; (3) learning of lemmas to
prove otherwise non-inductive properties; (4) explicit parallelism
for performance; and (5) removing weaker implied results
for human consumption.

\iffalse 
\subsubsection*{Example} To verify candidate loop invariants generated by
our algorithm at location $L$ in \lt{sqrt}, we first use backward analysis
to create the formulas $I: (x\ge 0 \wedge a = 0 \wedge s = 1, t=1)_0$
describing the program state when entering the loop the first time and
$T_n: s_{n-1}\le x_{n-1} \wedge x_n=x_{n-1} \wedge a_n= a_{n-1}+1 \wedge
s_{n}= s_{n-1}+t_{n-1}+2 \wedge t= t_{n-1}+2$ describing the state
transition upon re-entering the loop.

Next, we apply KIP on $I,T$ to check each candidate invariant iteratively.
KIP proves $t=2a+1$ with $k=0$ because both formulas $I \Rightarrow
(t=2a+1)_0$ and $(t=2a+1)_0 \wedge T_1 \Rightarrow (t=2a+1)_1$, representing
the base case and induction step when $k=0$, hold.  
Similarly, KIP proves $4s =
t^2 + 2t + 1$ is also $0$-inductive.  KIP proves  $s = (a+1)^2$  with $k=1$
because $(s=(a+1)^2)_0 \wedge T_1 \Rightarrow (s = (a+1)^2)_1$ fails (i.e.,
it is not $0$-inductive), but
$(s=(a+1)^2)_0 \wedge (s=(a+1)^2)_1 \wedge T_1 \wedge T_2 \Rightarrow (s =
(a+1)^2)_2$ holds.  The prover cannot show $s \ge t$ for $k \le 
\mathsf{maxK}$ because that invariant of \lt{sqrt} is not $k$-inductive
for any $k$.  However, after strengthening the induction step with
other proven results and iterating, KIP easily proves $s\ge t$. 
That is, $(t=2a+1, s = (a+1)^2 \wedge s \ge t)_0 \wedge T_1 \Rightarrow (s\ge
t)_1$ holds.  KIP identifies and rejects spurious results such as  $x\le
9989$ because the base case $I \Rightarrow (x\le 9989)_0$ does not hold
(e.g., with $x_0=1000$).
\fi 

\section{Experimental Evaluation}\label{results}

This section evaluates the efficiency and expressive power
of our methods. We consider the research questions:
  \begin{itemize}
  \item{ RQ1: Can the hybrid algorithm efficiently generate powerful 
  \emph{disjunctive} invariants and prove them correct? }
  \item{ RQ2: Is the hybrid algorithm effective on
  \emph{complex} correctness properties, such as those that
  are not classically inductive or involve nonlinear arithmetic? }
  \end{itemize}
To investigate RQ1 we applied our algorithms to a Disjunctive Invariant 
benchmark suite of kernels involving abstractions of string and array
processing.
To investigate RQ2 we used a Nonlinear Arithmetic benchmark suite. 
Each program comes equipped with ``gold standard'' full-correctness
annotations (e.g., assertions, postconditions, or formalized documented
invariants).

Each program was run on 300 random inputs to provide traces for invariant
generation and 100 random inputs for filtering, as described
in~\cite{nguyen:icse12:dig}. 
For small kernels, this yields sufficient
traces to generate accurate
invariants~\cite{sharma2013verification,NimmerE02:ISSTA}.
Our test programs come with annotated invariants at various locations such as loop heads and function exits. 
For evaluation purpose, we instrumented the values of variables at those locations and find invariants among the resulting traces.
We use only the weak max and min forms given in
Section~\ref{weak_maxplus} unless the number of variables is three 
or less, in which case it is also practical to use the general forms.

We implemented our algorithms in the dynamic analysis framework DIG~\cite{nguyen:icse12:dig,nguyen:tosem14:dig} using the Sage mathematical
environment~\cite{sage}. Our prototype uses the Tropical Polyhedra Library TPLib~\cite{allamigeon2013minimal} to manipulate  
max and min polyhedra and uses built-in Sage functions to solve equations
and construct convex hulls for classical polyhedra. The prototype KIP
prover uses Z3~\cite{demoura:tacas08:z3} to check the satisfiability of SMT formulas.  
As mentioned, we consider linear max-plus relations and set $\mathsf{maxK}=5$ by default.
The prototype constructs the verification conditions corresponding to
$M=(I,T)$  (Section~\ref{verify_with_KIP}) directly; a more
efficient tool such as Microsoft Boogie could also be used. 
The experiments were performed on a 64-core 2.60GHz Intel
Linux system with 128 GB of RAM; KIP used 64 threads of parallelism.

\begin{table}
\caption{Disjunctive Invariant experimental results.}  
\small
\centering
\begin{tabular}{l|c@{\ \ }c@{\ \ }r@{\ \ }r|rr|c@{}}
\textbf{Prog}& \textbf{Loc} &\textbf{Var} &\textbf{Gen} &\textbf{T$_\text{Gen}$} &
\textbf{Val} & \textbf{T$_{\text{Val}}$}& \textbf{Hoare}\\
\midrule
ex1      &1 &2 &  15 & 0.2     &4 &1.5      & \checkmark\\
strncpy  &1 &3 &  69 & 1.1     &4 &7.7      & \checkmark\\
oddeven3 &1 &6 &  286&  3.7   &8 &16.0     & \checkmark\\
oddeven4 &1 &8 &  867&  12.7  &22&46.0     & \checkmark\\
oddeven5 &1 &10& 2334&  56.8 &52&1319.4   & \checkmark\\
bubble3  &1 &6 &  249&   4.1   &8 &4.9      & \checkmark\\
bubble4  &1 &8 &  832&  11.7   &22&47.6     & \checkmark\\
bubble5  &1 &10& 2198&  53.9  &52&938.2    & \checkmark\\ 
partd3   &4 &5 & 479 &  10.5   &10&50.8     & \checkmark\\
partd4   &5 &6 & 1217&  23.3   &15&181.1    & \checkmark\\
partd5   &6 &7 & 2943&  53.3   &21&418.1    & \checkmark\\
parti3   &4 &5 &  464&  10.3  &10&45.5     & \checkmark\\
parti4   &5 &6 & 1148&  22.4  &15&165.1    & \checkmark\\
parti5   &6 &7 & 2954& 53.6   &21&405.6    & \checkmark\\
\bottomrule
\textbf{total} & & & 16055 & 317.6 & 264 & 3647.5 & 14/14 %Vu: remove 89 vars 10-10
\end{tabular}
\label{results1}
\end{table}

\iffalse
\begin{table*}
\small
\centering
\begin{tabular}{lc|crr|cccr|c}
\toprule
\textbf{Program}&\textbf{Locs} &\textbf{Invs} &\textbf{V} &\textbf{T$_\text{DIG}$}&\textbf{P, R, D (L)}&\textbf{U, E}  &$|k>0|$  & \textbf{T$_{\text{KIP}}$}&Verified\\
\midrule
ex1       &1 &  15 & 2&0.2     &4 , 6  ,5   ,6&0 , 0 &0, 0&1.5      &Y\\%VU: check this
strncpy   &1 &  69 & 3&1.1     &4 , 6  ,43  ,2&16, 0 &3, 1&7.7      &Y\\%VU: find out what this is
oddeven3  &1 &  286& 6 & 3.7   &8 , 77 ,201 ,0&0 , 0 &0, 0&16.0     &Y\\%VU: are these sorted ? 
oddeven4  &1 &  867& 8 & 12.7  &22, 206,639 ,0&0 , 0 &0, 0&46.0     &Y\\
oddeven5  &1 & 2334& 10 & 56.8 &52, 510,1772,0&0 , 0 &0, 0&1319.4   &Y\\
bubble3   &1 &  249& 6&  4.1   &8 , 76 ,165 ,0&0 , 0 &0, 0&4.9      &Y\\
bubble4   &1 &  832& 8& 11.7   &22, 203,607 ,0&0 , 0 &0, 0&47.6     &Y\\
bubble5   &1 & 2198& 10& 53.9  &52, 509,1637,0&0 , 0 &0, 0&938.2    &Y\\ 
partd3    &4 & 479 & 5& 10.5   &10, 96 ,363 ,0&10, 0 &0, 0&50.8     &Y\\
partd4    &5 & 1217& 6& 23.3   &15, 238,946 ,0&18, 0 &0, 0&181.1    &Y\\
partd5    &6 & 2943& 7& 53.3   &21, 485,2380,0&57, 0 &0, 0&418.1    &Y\\
parti3    &4 &  464& 5 & 10.3  &10, 98 ,351 ,0&5 , 0 &0, 0&45.5     &Y\\
parti4    &5 & 1148& 6 & 22.4  &15, 214,895 ,0&24, 0 &0, 0&165.1    &Y\\
parti5    &6 & 2954& 7 &53.6   &21, 456,2406,0&71, 0 &0, 0&405.6    &Y\\
mcarthy91 &1 &     &   &       &              &      &    &         &N\\
\bottomrule
\end{tabular}
\caption{Experimental Results 1.}
\label{results1}
\end{table*}
\fi

\subsection{RQ1: Disjunctive Invariants}

We evaluate our approach on several benchmark kernels for disjunctive
invariant analysis~\cite{allamigeon2008inferring}, listed in
Table~\ref{results1}. These programs typically have many execution paths, 
e.g., \lt{oddeven5} contains 12 serial ``if'' blocks and thus $2^{12}$
paths. The documented correctness assertions for these
programs require reasoning about disjunctive invariants,\footnote{We note
that this suite is not exhaustive. Max-plus algebra is still relatively
new, and while it has real-world applications such as network traffic
shaping~\cite{daniel2006max, heidergott2006max} and biological sequence
alignment~\cite{comet2003application}, to our knowledge this is the first
paper on dynamic inference for max-plus invariants and thus few benchmarks
are yet available.}
but do not involve higher-order logic. For example, the sorting procedures
are asserted to produce sorted output, but are not asserted to produce a
permutation of the input.

Table~\ref{results1} report experimental results.
The \textbf{Loc} column lists the number of locations where invariants
were generated.
The \textbf{Var} column reports the number of distinct variables involved
in the invariants. The \textbf{Gen} column counts the number of unique
candidate invariants generated by our dynamic algorithm. The
\textbf{T$_\text{Gen}$} column reports the generation and filtering time,
in seconds, averaged over five runs. 
The number of generated invariants speaks to the expressive power of
the algorithm: higher is better, indicating that we can reason about
more disjunctive relationships over program variables. 
Time indicates the efficiency of our algorithm: lower is better. 
The \textbf{Val} column reports the number of generated invariants that
KIP proved correct and non-redundant with respect to the program. 
The other generated invariants were disproved three times as often
as they were proved redundant.
A few invariants, just under 2\% on
average, could neither be proved nor disproved. The \textbf{T$_\text{Val}$}
column counts the time, in seconds, to analyze all of the generated
invariants. 

We desire validated invariants to statically prove each program's
annotated correctness condition via Hoare logic. The \textbf{Hoare} column
indicates whether the validated invariants were sufficient to
prove program correctness. For all of these programs, the invariants
generated and validated by our hybrid approach---an average of 18 per
programs---were sufficient for a static proof of full correctness. 

For example, for the C string function \lt{strncpy}, which copies the first
$n$ characters from a (null-terminated) source $s$ to a (unconstrained)
destination $d$, we inferred the relation:
\[
\small
(n \ge |s| ~\wedge~ |d|=|s|) ~\vee~ (n < |s| ~\wedge~ |d| \ge n)
\]
This captures the desired semantics of the function: 
if $n \ge |s|$, then the copy stops at the null terminator of
$s$, which is also copied to $d$, so $d$ ends up with the same length as $s$.
However, if $n < |s|$, 
then the terminator is not copied to $d$, so 
$|d| \ge n$. 

As a second example, for \lt{bubble}$_N$ and \lt{oddeven}$_N$, which sort the
input elements
$x_0,\dots,x_N$ and store the results in
$y_0,\dots,y_N$, our inferred invariants
prove the outputs $y_0$ and $y_N$ hold the smallest and largest elements of
the input.  However, we cannot show that $y$ is a
permutation of $x$ because that is expressible only under
higher-order logics (our results here are similar to those 
of purely static analyses~\cite{allamigeon2008inferring}).

\iffalse
Eight of the invariants were not $\mathsf{maxK}$-inductive but were
still validated because of our technique's iterative re-use of learned
invariants. Four of the invariants we validated were not classically
inductive and thus required $k$-induction (see the next subsection for more
such examples).
\fi 

Table~\ref{results1} shows that our method is efficient. 
We can infer about 3000 disjunctive relations per minute, on average, and
validate about 300 per minute. The method is also effective. 
We produced 264 non-redundant, proved-correct disjunctive invariants, and those
invariants were sufficient to statically prove each program's contract.

\subsection{RQ2: Complex Invariants}
\label{sec-rq2} 

\begin{table}
\caption{Nonlinear Arithmetic experimental results.}
\small
\centering
\begin{tabular}{l|c@{\ \ }c@{\ \ }r@{\ \ }r|r@{\ }r@{\ }r|c@{}}
\textbf{Prog} & \textbf{Loc}&\textbf{Var} & \textbf{Gen} &\textbf{T$_\text{Gen}$}& 
\textbf{Val} & \textbf{$k$I}& \textbf{T$_{\text{Val}}$}& 
\textbf{Hoare}\\
\midrule
cohendv & 2 & 6& 152 & 26.2   & 7 &14& 8.2  & \checkmark \\ %7+7+135+3+1 = 152
divbin  & 2 & 5& 96   & 37.7  &  8&15& 8.7   & -- \\ %requires is_even(b)
manna   & 1 & 5& 49   & 19.2  &  3&2 & 5.6   & \checkmark \\  
hard    & 2 & 6& 107  & 14.2  & 11&4 & 9.2   & -- \\ %1 z3 unknown,requires is_even(d),is_even(p)
sqrt1   & 1 & 4& 27   & 25.3  &  3&1 & 4.3   & \checkmark \\ %26(2) but with the annotated inv (so that it's consistent w/ paper)
dijkstra& 2 & 5& 61  & 30.7   &  8&6 & 10.9  & -- \\ %1 3 MANUALLY IGNORED
freire1 & 1 & 3& 25   & 22.5  &  2&0 & 2.2   & \checkmark \\ 
freire2 & 1 & 4& 35   & 26.0  &  3&1 & 5.1   & \checkmark \\  %TODO: check me
cohencb & 1 & 5& 31   & 23.6  &  4&1 & 4.2   & \checkmark \\ 
egcd1   & 1 & 8& 108  & 43.1  &  1&8 & 12.8  & -- \\ %P2 requires P3,P3 requires P2,3 froze
egcd2   & 2 &10& 209  &60.8   &  8&12& 14.6  & \checkmark \\ 
egcd3   & 3 &12& 475  &67.0   & 14&25& 23.4  & \checkmark \\
lcm1    & 3 & 6& 203  & 38.9  & 12&0 & 14.2  & \checkmark \\ 
lcm2    & 1 & 6& 52   & 14.9  &  1&10& 0.9   & \checkmark \\ 
prodbin & 1 & 5& 61   & 28.3  &  3&10& 1.1   & -- \\ %1 z3 unknown
prod4br & 1 & 6& 42   & 9.6   &  4&7 & 8.6   & \checkmark \\ 
fermat1 & 3 & 5& 217  &75.7   &  6&1 & 6.2   & \checkmark \\
fermat2 & 1 & 5& 70   & 25.8  &  2&0 & 5.2   & \checkmark \\
knuth   & 1 & 8& 113  & 57.1  &  4&6 & 24.6  & \checkmark \\ 
geo1    & 1 & 4& 25   & 16.7  &  2&4 & 1.5   & \checkmark \\ 
geo2    & 1 & 4& 45   & 24.1  &  1&10& 2.1   & \checkmark \\ 
geo3    & 1 & 5& 65   & 22.1  &  1&12& 2.7   & \checkmark \\ 
ps2     & 1 & 3& 25   & 21.1  &  2& 0& 4.0   & \checkmark \\ 
ps3     & 1 & 3& 25   & 21.9  &  2& 0& 4.2   & \checkmark \\ 
ps4     & 1 & 3& 25   & 23.5  &  2& 0& 4.9   & \checkmark \\ 
ps5     & 1 & 3& 24   & 24.9  &  2& 0& 7.4   & \checkmark \\ 
ps6     & 1 & 3& 25   & 25.0  &  2& 0& 69.5  & \checkmark \\  %3 z3 freezes  %ps6 which uses nonlinear degree up to 6 has the most error
\bottomrule
\textbf{total} & & & 2392 & 825.9 & 118 & 149 & 266.3 & 22/27 %remove 142 vars
\end{tabular}
\label{results2}
\end{table}

We also evaluate our technique on more complex programs, such as those 
that are not classically inductive or use nonlinear arithmetic, 
by studying the NLA (nonlinear arithmetic) test
suite~\cite{nguyen:icse12:dig}.  The suite consists of 27 programs from various
sources collected by Rodr\'iguez-Carbonell and
Kapur~\cite{1236086,1222597,RCEthesis}.  
The programs are relatively small, on average two loops of 20 lines of
code each.  However, they implement nontrivial mathematical
algorithms and are often used to benchmark static analysis methods.  
For these programs, we generate and check loop invariants of two polynomial
forms: nonlinear equations and linear max-plus inequalities among program
variables.  
We consider at most 200 generated terms per polynomial equality, e.g., invariants up to degree five if four variables are involved.
The documented correctness assertions for these 27 programs
require nonlinear invariants, mostly equalities among nonlinear polynomials.

Table~\ref{results2} shows the results, in a format similar to that
of Table~\ref{results1}. The large number of candidate
invariants generated---over 80 per program, on average---highlights the
expressive power of our technique. The generation is slightly slower
than for the disjunctive benchmarks because these require equation
solving for large numbers of terms representing nonlinear polynomials.
However, our weak forms take an order of magnitude less time than do the general
equality relations. The overall generation process remains efficient,
averaging thirty seconds per program. 

Our hybrid approach is able to formally validate 118 of those invariants,
or 4.3 per program on average, proving them correct and non-redundant. 
The validation is rapid (0.1 seconds per candidate invariant, on average,
compared to 0.2 for the disjunctive benchmarks) but here shows its
reliance on the underlying SMT theorem prover. 
For 18 of these 27 programs, some of the theorem prover queries
issued caused the Z3 SMT solver to return an
\emph{unknown} error or stop responding. This is likely due to the recent
addition of support for nonlinear arithmetic, and we reported these
errors to the Z3 developers. In the interim, however, such candidate
invariants must be rejected. 

The \textbf{$k$I} column in Table~\ref{results2} counts the number of
invariants that require $k$-induction to be proved or disproved. 
Similarly, an additional 39 of the proved invariants required 
considering discovered invariants as lemmas, and were
not otherwise $\mathsf{maxK}$-inductive. The significant presence of
invariants requiring $k$-induction or learned lemmas validates the
KIP architecture design choice. 

Ultimately, the invariants generated and validated by our technique can be
used to statically prove the correctness of 22 of these 27 programs using
Hoarse logic. Of the remainder, two require novel invariant forms,
one requires invariants that are not
$k$-inductive, and two are correct but beyond our
current SMT solver.  For the first type, \lt{divbin} requires
the invariant $\exists k. x=2^k$, and our algorithm does not support 
exponential forms. The \lt{hard} program also has similar exponential invariants. For the 
second type, our dynamic algorithm generates three non-linear
equalities that precisely capture \lt{egcd1}'s semantics, and manual
inspection verifies that they are not $k$-inductive for any $k$, and thus KIP
cannot prove them. 
For the third type, 
our dynamic algorithm generates invariants that precisely
capture the semantics of \lt{prodbin} and \lt{dijkstra} 
and KIP can process them, but the backend SMT
solver hangs instead of proving them (we have manually verified that they
are otherwise correct). Thus we could prove two more programs with
a better SMT solver, two more programs 
with a better theorem prover architecture, and could not prove
the last without a new algorithm for invariant generation.

\section{Related Work}\label{related_work}

{\bf Dynamic invariant analyses.} 
Daikon~\cite{daikon} is a popular and influential dynamic invariant
analysis that infers candidate invariants using templates.  Daikon
comes with a large list of invariant templates and returns those 
that hold over a set of program traces.  Daikon can use ``splitting''
conditions~\cite{ernstcollections} to find disjunctive invariants
such as ``$\mathsf{if}\;c \; \mathsf{then}\; a\; \mathsf{else}\; b$''.  Our
algorithm does not depend on splitting conditions and our max- and
min-plus disjunctive invariants are more expressive than those
currently supported by Daikon.

Recently, Sharma \emph{et al.}~\cite{sharma2013verification} proposed a
machine-learning based approach to find disjunctive invariants.
Their method operates on traces representing good and bad
program states: good traces are obtained by running the program on random
inputs and bad traces correspond to runs on which an assertion or
postcondition is violated. They use a \emph{probably approximately correct}
machine learning model to find a predicate, representing a candidate
program invariant, that separates the good and bad traces. For efficiency,
they restrict attention to the octagon domain and search only for
predicates that are arbitrary boolean combinations of octagon inequalities.
Finally, they use standard induction technique to check the candidate
invariants using Z3~\cite{demoura:tacas08:z3}. While our method shares their focus on
disjunctive invariants, a key difference is that the strength of their
results depends strictly on existing annotated program assertions. For
example, in \lt{ex1}, if the line \lt{assert(y==11)} is not provided by
the programmer then their method will only produce the trivial invariant
$\mathsf{True}$. By contrast, our approach does not make such assumptions
about the input program, and in some sense the purpose of our approach is
to generate those assertions.

{\bf Hybrid approaches.} 
Nimmer and Ernst integrated the ESC/Java static
checker framework~\cite{Nimmer01staticverification} with Daikon, allowing them to
validate candidate invariants using a Hoare logic verification approach.
This work is very similar in motivation and architecture to ours. Key
differences include our detection of richer disjunctive invariants, our
verification with respect to full program correctness (``Rather than
proving complete program correctness, ESC detects only certain types of
errors''~\cite[Sec.~2]{Nimmer01staticverification}), and our larger
evaluation (our system proves over four times as many non-redundant
invariants valid and considers over four times as many benchmark kernels). 

{\bf Static max-plus analyses.} 
The static analysis work of Allamigeon \emph{et al.}~\cite{allamigeon2008inferring} uses abstract interpretation to
approximate program properties under the max- and min-plus domains. In
contrast to our work, which computes max-plus formulas from dynamic traces,
their method starts directly from a formula representing an initial
approximation of the program state space and gradually improves that
approximation based on the program structure until a fixed point is
reached. 
As with other abstract interpretation approaches for inferring disjunctive invariants such as~\cite{sankaranarayanan2006static,popeea2007inferring}, 
their method uses an ad-hoc widening operator to ensure termination.

The recent static analysis work of Kapur \emph{et
al.}~\cite{kapur:arm13:geometric} uses quantifier elimination to find max
invariants over pairs of variables. Their method uses table look-ups to
modify max relations based on the program structures (e.g., to
determine how the max relation is changed after an assignment $a=a+10$).
For scalability the approach restricts attention to specific program
constructs. For example, they only support analysis on assignments or
guards that do not involve multiplication. 

A high-level difference between such techniques and our work is that
we focus on the efficient inference of invariants from dynamic traces. 
More generally, we hypothesize that the weak max and min forms introduced
in this paper would allow such static techniques to be practically applied
to more general classes of programs. 

{\bf Uses of $k$-induction.} 
The application of $k$-induction is becoming increasingly popular 
for formulas that may not admit classic induction.
Sheeran \emph{et al.} applied $k$-induction to verify hardware designs
using SAT solvers~\cite{sheeran2000checking}.  The PKIND model checker of
Kahsai and Tinelli~\cite{kahsai2011instantiation, KahTin-PDMC-11} uses $k$-induction and SAT/SMT solvers to verify synchronous programs in the Lustre language. Recently, Donaldson
\emph{et al.}~\cite{donaldson11} applied $k$-induction to imperative
programs with multiple loops. A key distinction between our KIP
architecture and these approaches is that none of them offers all four of
the other properties (SMT, lemma re-use, redundancy elimination and
parallelism) that we find critical for efficiently verifying large numbers
of candidate invariants over programs with complex properties such as
nonlinear arithmetic. However, we note that the programs and candidate
invariants learned in this paper could serve as a benchmark suite
for the evaluation of such theorem provers (i.e., hundreds of valid and
invalid formulas involving nonlinear arithmetic, many of which are
$k$-inductive).

\section{Conclusion}\label{conclusion}

Program invariants are important for defect detection, program
verification, and automated repair. Existing approaches struggle with 
soundness and expressive power and cannot learn
disjunctive invariants. We propose a hybrid approach to invariant
inference that finds complex invariants dynamically and proves them
statically. 

We present the first dynamic algorithm to learn the max-plus class of
disjunctive invariants, allowing us to capture conditional
behavior. To do so, we reformulate the problem of convex invariant
detection in a non-standard max-plus algebra. We gain expressive power
with dual min-plus constraints, capturing if-and-only-if
behavior. 

Critically, we also define and infer a new class of 
\emph{weak} max- and min-plus
invariants that retain useful expressive power while requiring only
polynomial complexity. To the best of our knowledge, this is the first 
use of a restricted form of max- or min-plus invariants. 

These weak forms suggest new
theoretical research directions for max- and min-plus algebras. Although
we provide the algorithm for computing weak relations given
points, the dual problem for computing extremal points given relations
remains open. The cost of computing general polyhedra inspired
general research work on weaker abstract domains (e.g., interval, box, zone
and octagon) and formal logic~\cite{mine2004weakly}. We see the
relationship between our weak max-plus form and general max-plus as
analogous to that between octagons and general polyhedra (e.g., fixed
coefficients, only one open parameter, etc.), and hope that other max-plus
researchers may find our weak form somewhat as useful as polyhedra
practitioners have found octagon forms.

We also propose a static approach to invariant verification based on
iterative, parallel $k$-inductive SMT theorem proving. Many program
invariants are not classically inductive, and 
$k$-induction allows us to prove them. Similarly, the re-use of
learned invariants as lemmas allows our system to prove 
non-$\mathsf{maxK}$-inductive invariants in practice. 
Our design's explicit parallel structure is critical
for performance. By construction, our algorithm never overapproximates if
the real invariant is expressible in our system, and
validating each candidate against the program means that our system never
underapproximates: this approach helps address the issue of spurious or
incorrect invariants.

We evaluate our algorithm by extending the DIG framework and considering
difficult benchmark kernels involving nonlinear arithmetic and
abstract arrays.  Our approach is efficient and effective at finding
and validating disjunctive, non-linear and complex invariants. 
Ultimately we find and verify
invariants that are powerful enough to prove 36 of 41 programs correct
using Hoare logic, taking two minutes per program, on average, and 
producing no spurious answers.

\section{ACKNOWLEDGMENTS} 
\begin{sloppypar}
Nguyen is grateful for an internship at the Naval Research Laboratory, which introduced him to $k$-induction proving and led to KIP.
We thank Matthias Horbach and Hengjun Zhao for insightful discussions.
We gratefully acknowledge the partial support of AFORSR (FA9550-07-1-0532, FA9550-10-1-0277), DARPA (P-1070-113237), DOE (DE-AC02-05CH11231), NSF (SHF-0905236, CCF-0729097, CNS-0905222, CCF-1248069, DMS-1217054) and the Santa Fe Institute.
\end{sloppypar}

\balance
\bibliographystyle{abbrv}
\bibliography{mpp}
\end{document}